\documentclass{llncs}
\usepackage{amssymb} %lettres mathématiques

\let\emptyset\varnothing
\usepackage{cite}
\usepackage{booktabs}
\usepackage{url}
\usepackage{xspace}
\usepackage{tikz}
\usetikzlibrary{shapes,arrows}
\usepackage{pgfplots}
\usepackage{float}
\restylefloat{table}

\tikzstyle{line}=[draw]
\tikzstyle{cloud} = [draw, ellipse,fill=white, node distance=3cm,
    minimum height=2em]
\def\holfour{\textsf{HOL4}\xspace}
\def\isabellehol{\textsf{Isabelle/HOL}\xspace}
\def\hollight{\textsf{HOL Light}\xspace}
\def\coq{\textsf{Coq}\xspace}

\newcommand{\ra}[1]{\renewcommand{\arraystretch}{#1}}

\begin{document}

\author{Thibault Gauthier \and Cezary Kaliszyk}
\title{Matching concepts across HOL libraries \thanks{The final publication is available at http://link.springer.com.}}

\date{\today}
\institute{University of Innsbruck, Austria\\
\{thibault.gauthier,cezary.kaliszyk\}@uibk.ac.at}
\maketitle{}

\begin{abstract}
  Many proof assistant libraries contain formalizations of the same
  mathematical concepts. The concepts are often introduced (defined)
  in different ways, but the properties that they have, and are in
  turn formalized, are the same. For the basic concepts, like natural numbers,
  matching them between libraries is often straightforward, because of
  mathematical naming conventions. However, for more advanced concepts,
  finding similar formalizations in different libraries is a non-trivial
  task even for an expert.

  In this paper we investigate automatic discovery of similar concepts
  across libraries of proof assistants. We propose an approach for
  normalizing properties of concepts in formal libraries and a number of
  similarity measures. We evaluate the approach on HOL based proof assistants
  \holfour, \hollight and \isabellehol, discovering 398 pairs
  of isomorphic constants and types.
\end{abstract}

\section{Introduction}
%\subsection{Motivation}
Large parts of mathematical knowledge formalized in various
theorem provers correspond to the same informal concepts. Basic
structures, like integers, are often formalized not only in
different systems, but sometimes also multiple times in the same
system. There are many possible reasons for this: the user may for
example want to investigate special features available only for certain
representations (like code extraction~\cite{haftmann-itp13}), or simply
check if the formal proofs can be done in a more straightforward manner
with the help of alternate definitions. 
With multiple proof assistants, even the definitions of basic concepts
may be significantly different: in \isabellehol~\cite{isabelle}
the integers are defined
as a quotient of pairs of naturals, while in \hollight~\cite{Harrison09hollight} they are a subset
of the real numbers. Typically the proofs concerning a mathematical concept formalized in one
system are not directly usable in the other, so a re-formalization
is necessary.

The idea of exchanging formal developments between systems has been
investigated both theoretically and practically many times~\cite{obuaimport,holcoq,ckak-itp13}. Typically
when a concept from the source systems is translated to a target system,
and the same concept exists in the target system already, a new isomorphic
structure is created and the relation between the two is lost. The properties
that the two admit are the same and it is likely that the user formalized many
similar ones.

%\subsection{Approach}

In this work we investigate automatic discovery of such isomorphic
structures mostly in the context of higher order logic. Specifically
the contributions of this work are:

\begin{itemize}
\item We define patterns and properties of concepts in a formal library
  and export the data about constants and types from \hollight, \holfour,
  and \isabellehol together with the patterns.
\item We investigate various scoring functions for automatic discovery
  of the same concepts in a library and across formal libraries and evaluate
  their performance.
\item We find 398 maps between types and constants of the three
  libraries and show statistics about the same theorems in the libraries,
  together with normalization of the shape of theorems.
\end{itemize}

%\subsection{Related work}
There exists a number of translations between formal libraries.
The first translation of proofs that introduced maps between concepts was the
one of Obua and Skalberg~\cite{obuaimport}.  There, two commands for
mapping constructs were introduced: \texttt{type-maps} and
\texttt{const-maps} that let a user map \hollight and \holfour
concepts to corresponding ones in \isabellehol.  Given a type (or
constant) in the maps, during the import of a theorem all occurrences
of this type in the source system are replaced by the given type of
the target system. In order for this construction to work, the basic
properties of the concepts must already exist in the target system,
and their translation must be avoided. Due to the complexity of
finding such existing concepts and specifying the theorems which do
not need to be translated, Obua and Skalberg were able to map only
small number of concepts like booleans and natural numbers, leaving
integers or real numbers as future work.

The first translation that mapped concepts of significantly different
systems was the one of Keller and Werner~\cite{holcoq}. The translation
from \hollight to \coq proceeds in two phases. First, the HOL proofs
are imported as a defined structures. Second, thanks to the \emph{reflection}
mechanism, native \coq properties are built. It is the second phase that
allows mapping the HOL concepts like natural numbers to the \coq standard
library type \texttt{N}.

The translation that maps so far the biggest number of concepts has
been done by the second author~\cite{ckak-itp13}. The translation process
consists of three phases, an exporting phase, offline processing and
an import phase. The offline processing provides a verification of the
(manually defined) set of maps and checks that all the needed theorems
will be either skipped or mapped. This allows to quickly add mappings
without the expensive step of performing the actual proof translation,
and in turn allows
for mapping 70 \hollight concepts to their corresponding \isabellehol counterparts.
All such maps had to be provided manually.

Bortin et al.~\cite{bortin06awe} implemented the AWE framework which allows
the reuse of \isabellehol formalization recorded as a proof trace multiple
times for different concepts. Theory morphisms and parametrization are
added to a theorem prover creating objects with similar properties. The use
of theory morphisms together with concept mappings is one of the basic
features of the MMT framework~\cite{rabe13mmt}. This allows for mapping
concepts and theorems between theories also in different logics. So far
all the mappings have been done completely manually.

Hurd's OpenTheory~\cite{opentheory} aims to share specifications and proofs
between different HOL systems by defining small theory packages. In order to
write and read such theory packages by theorem prover implementations
a fixed set of concepts is defined that each prover can map to. This provides
highest quality standard among the HOL systems, however since the
procedure requires manual modifications to the sources and inspection of
the libraries in order to find the mappings, so far only a small number of
constants and types could be mapped. Similar aims are shared by semi-formal
standardizations of mathematics, for example in the OpenMath content dictionaries.
For a translation between semi-formal mathematical representation again concept
lookup tables are constructed manually~\cite{watt06,davenport01}.

The proof advice systems for interactive theorem proving have studied similar
concepts using various similarity measures. The methods have so far been mostly
restricted to similarity of theorems and definitions. They have also been
limited to single prover libraries. Heras and Komendantskaya in the proof
pattern work~\cite{jhek14pattern} try to find similar Coq/SSReflect definitions
using machine learning. Hashing of definitions in order to discover constants
with same definitions in Flyspeck has been done in~\cite{KaliszykUrban14msc}.
Using subsumption in order to find duplicate lemmas has been explored in the
MoMM system~\cite{Urban06momm} and applied to \hollight lemmas in~\cite{ckju-lpar13}.

The rest of this paper is organized as follows: in Section~\ref{s:export} we describe
the process of exporting the concepts like types and constants from three provers. In Section~\ref{s:classify}
we discuss the classification of patterns together with the normalization of theorems, while in
Section~\ref{s:crosslib} we define the scoring functions and an iterative matching algorithm.% tested in our experiments. 
We present the results of the experiments in Section~\ref{s:experiment} and
in Section~\ref{s:concl} we conclude and present an outlook on the future work.

\section{The theorem and constant data}~\label{s:export}
In this section we shortly describe the data that we will perform our experiments on and
the way the theorems and constants are normalized and exported. We chose three proof
assistants based on higher-order logic: \holfour~\cite{hol4}, \hollight~\cite{Harrison09hollight}
and \isabellehol~\cite{isabelle}. The sizes of the core libraries of the three are
significantly different, so in order to get more meaningful results we export library
parts of the same order of magnitude. This amounts to all the theories included with the
standard distribution of \holfour. In case of \hollight
we include multivariate analysis~\cite{multivariate}, HOL in HOL~\cite{holinhol} and the
67 files that include the proofs of the 100 theorems~\cite{17provers} compatible with the
two. For Isabelle we export the theory \texttt{Main}.

%We will describe thoroughly how the exporting phase was programmed in \holfour, and then explain how export is done differently in \isabellehol and \hollight. The \holfour proof assistant is organized as a library of dependent theories. All these theories are manually listed using the HOL Reference Page. Each theory is loaded and followed by a call to 
The way to access all the theorems and constants in \hollight has been described in
detail in~\cite{KaliszykUrban14jar} and for \holfour and \isabellehol accessing values
of theories can be performed using the modules provided by the provers (\texttt{DB.thms}
and \texttt{@\{theory\}} object respectively). We first perform a minimal normalization of the
forms of theorems (a further normalization will be performed on the common representation
in Section~\ref{s:classify}) and export the data. We will focus on \holfour, the procedures in the other two are similar.

The hypotheses of the theorems are discharged and all free variables are generalized.
In order to avoid patterns arising from known equal constants, all theorems of the
form $\vdash c_1 = c_2$ (in \holfour four of them are found by calling \texttt{DB.match})
are used to substitute $c_1$ by $c_2$ in all theorems.

The named theorems and constants are prefixed with
theory names and explicit category classifiers (\texttt{c} for constants, \texttt{t} for theorems)
to avoid ambiguities. Similarly, variables are explicitly numbered with their position of the
binding $\lambda$ (this is equivalent to the de Bruijn notation, but possible within the
data structure used by each of the three implementations). We decided to include the type information
only at the constant level, and to skip it inside the formulas.
  \begin{example}
  $\forall x:int. \ x = x \longrightarrow cHOL4.bool.\forall \ (\lambda V.((cHOL4.min.\!=\  V)\ V)) $
  \end{example}
Analogously, for all the constants their most general types are exported. Type variables are
normalized using numbers that describe their position and type constructors are prefixed
using theory identifiers and an explicit type constructor classifier.
  \begin{example}
  $(num,a) \longrightarrow  tHOL4.pair.prod(tHOL4.num.num,Aa) $
  \end{example}
The numbers of exported theorems and constants are presented in Table~\ref{fig:thno}.
\begin{table}[thb]
  \centering\ra{1.3}
  \begin{tabular}{@{}lcccccc@{}}
    \toprule 
    & \phantom{abc} & 
    \hollight & \phantom{abc} & 
    \holfour  & \phantom{abc} & 
    \isabellehol \\
    \midrule
    Number of theorems && 11501 && 10847 && 18914  \\
    Number of constants && 871 && 1962 && 2214 \\
    \bottomrule
  \end{tabular}
  \caption{Number of theorems and constants after the exporting phase\label{fig:thno}} 
\end{table}

\section{Patterns and classification}\label{s:classify}

In this section we will look at the concept of \emph{patterns} created from
theorems, which is crucial in our classification of concepts and the
algorithms for deriving patterns and matching them. In the following we
will call the constants and types already mapped to concepts as \emph{defined}.
%\subsection{Definition of pattern}
\begin{definition}[pattern]
 Let $f$ be a formula with no free variables and $C$ the set of its constants. Let $D = \lbrace d_1,\ldots,d_n \rbrace$ be a set of defined constants and $A = C \setminus D = \lbrace a_1,\ldots,a_m \rbrace$ a set of undefined constants. Its pattern is defined by: 
   \[P (f[a_1,\ldots,a_m,d_1,\ldots,d_n]) := \lambda a_1 \ldots a_n. f[a_1,\ldots,a_n,d_1,\ldots,d_n] \] 
\end{definition}
  \begin{example} The pattern of $\forall x\ y.\ x * y = y * x$ is:\\
  - with $D = \lbrace \forall, = \rbrace$,
 $\ \  \lambda a_1. \ \forall x\ y.\ a_1\ x\ y = a_1\ y\ x$.\\
 - with $D = \lbrace \forall \rbrace$,
  $\ \ \lambda a_1 a_2. \ \forall x\ y.\ a_1 (a_2\ x\ y) (a_2\ y\ x)$.\\
 - with $D = \emptyset$,
 $\ \  \lambda a_1 a_2 a_3. \ a_1\ \lambda x\ y.\ a_2 (a_3\ x\ y) (a_3\ y\ x)$.
  \end{example}

  Patterns are equal when they are $\alpha$-equivalent.
  In practice, we order the variables and constants by the order in which they appear when traversing the formula from top to bottom. This means that checking if two formulas are $\alpha$-equivalent amounts
to verifying the equality of their patterns with no constants abstracted.

%\subsection{Creation of patterns and normalization}

The formulas exported from all proof assistant libraries are parsed to a standard representation ($\lambda$-terms).
The basic logical operators of the different provers are mapped to the set of defined constants
and the theorems are rewritten using these mappings before further normalization. Finally, the
patterns of the normalized formulas are extracted according to the specified defined constants.

We define three ways in which patterns are derived from the formula, each corresponding to a
certain level of normalization:

\paragraph{$norm_0$}: Given $D = \emptyset$ we can define a pattern
corresponding to the theorem without any abstraction (identity).

\paragraph{$norm_1$}: With $D = \lbrace \forall,\exists,\wedge,\vee,\Rightarrow,\neg,= \rbrace$ ($\Leftrightarrow$ is considered as $=$). The procedure is similar to the normalization done
by first order provers (to the conjunctive normal form) with the omission of transformations on existential
quantifiers, as we do not want do perform skolemization. We additionally normalize associative
and commutative operations. The procedure performs the following steps at every formula level:
        \begin{itemize}
        \item remove implication, 
        \item move negation in,
        \item move universal quantifiers out (existential quantifiers are not moved out to maximize the number of disjunctions in the last step),
        \item distribute disjunction over conjunctions,
        \item rewrite based on the associativity of $\forall,\exists,\wedge$ and $\vee$,
        \item rewrite based on the commutativity of $\forall,\exists,\wedge,\vee$ and $=$,
        \item separate disjunctions at the top formula level (example below).
        \end{itemize}
     \begin{example} $\forall x\ y.\ (x \geq 0 \wedge x \leq y) \longrightarrow (\forall x.\ x \geq 0) \wedge (\forall x\ y.\ x \leq y)$
     \end{example}

\paragraph{$norm_2$}: Aside from all the normalizations performed by $norm_1$, we
  additionally consider a given list of associative and commutative constants 
  (see Table~\ref{tab:prop1} in Section~\ref{s:experiment}) that is used to
  further normalize the formula. The set of defined constants stays the same as $norm_1$, which in particular means that the associative - commutative (AC) constants stay undefined and can be abstracted over.\\

\noindent Given the normalized theorems we will look at patterns relative to constants.
In the following, we will assume that the constants are partitioned in ones
that have been defined (mapped to a constant) and undefined.

%\subsection{Relative patterns}

\begin{definition}[pattern relative to a constant]\label{def:relative}
    Let $a_{i-1}$ be an undefined constant appearing in a formula $f$ in the $i^{th}$ position. The pattern of $f$ relative to $a_{i-1}$ is defined by:
      \[ P_{a_{i-1}}(f) := (P(f),i-1) \]
\end{definition}
\begin{example}
  Suppose $D = \emptyset$. Then the only two patterns that the reflexivity principle induces are:
  \begin{eqnarray*}
  P_{\forall} (\forall x.\ x = x) &=& (\lambda a_0 a_1.\ a_0 \ (\lambda v_0.\ a_1 \ v_0 \ v_0),0)\\
  P_{=} (\forall x.\ x = x) &=& (\lambda a_0 a_1.\ a_0 \ (\lambda v_0.\ a_1 \ v_0 \ v_0),1)\\
  \end{eqnarray*}
\end{example}
Typically, we will be interested in patterns where $D$ includes the predicate logic constants, so
the reflexivity principle will not produce any patterns. The patterns will be properties of
operations like commutativity or associativity. In order to find all such properties we define:

\begin{definition}
The set of patterns associated with a constant $c$ in a library $lib$ is defined by:
    \[ P^{set}(lib,c) = \bigcup \limits_{f \in lib} P_{c}(f) \]
Let $(abs,i)$ be a relative pattern. Its associated set of constants, in library $lib$, is:
    \[ C^{set}(lib,(abs,i)) := 
      \lbrace c \in lib , \exists f \in lib, \ P_c(f) = (abs,i)) \rbrace 
      \]

\end{definition}
We can now define one of the basic measures we will use for comparing similarity of constants:
\begin{definition}
The set of common
relative patterns shared by a constant $c_1$ in $lib_1$, and a constant $c_2$ in $lib_2$ is:
        \[ P^{set}(lib_1,c_1) \cap P^{set}(lib_2,c_2) \]
\end{definition}
%\begin{remark}
In the remaining part of this paper, we will not always specify if a pattern is relative
or not. %Since it is clear from the context 
%\end{remark}

%\subsection{Type patterns and derived type matches}  
We proceed with forming type patterns. Type patterns are defined in a similar way
to formula patterns. Types are partitioned into already defined types (initially the type of
booleans -- propositions) and undefined types. Type variables are also considered as undefined
to enable their instantiation, and the list of leaf and node types involved is saved to allow
matching.
\begin{example}
  Let $D^{type} = \lbrace fun \rbrace$ and $a$ be a type variable. Then:
  \begin{eqnarray*}
   & P^{type} ((a \rightarrow a,int \rightarrow int)) = P^{type} ((pair(fun (a,a),fun(int,int)))) &\\
   & = (\lambda a_0 a_1 a_2.\ (a_0(fun(a_1,a_1),fun(a_2,a_2))),[pair,a,int]) &\\
  \end{eqnarray*}
  \end{example}
   Suppose we are given two types with respective patterns $(abs_1,[t_1 \ldots t_n])$ and $(abs_2,[u_1 \ldots u_m])$. They match if $abs_1$ is $\alpha$-equivalent to $abs_2$. The list of their derived type matches is $[(t_1,u_1),\ldots,(t_n,u_n)]$, from which the pairs containing at least one type variable are removed.

\section{Matching concepts across libraries}\label{s:crosslib}
In this section, we will investigate measures of similarity in order to find
the same types and constants between libraries. First, we will define a similarity score
for each pair of constants. Then, we will suppose that the best match is correct
and use it to update the similarity scores of the other pairs iteratively.

\subsection{Similarity score}
The easiest way to tell if two constants are related is to look at the number
of patterns they share. However, the more a pattern has associated constants, the
less relevant it is. To test each of these possibilities, two weighting functions
are defined:
    \[ w_0(lib,p) = 1, \ \ \
       w_1(lib,p) = \frac {1} {card(C^{set}(lib,p))}  \]
where $p$ is a pattern in library $lib$. The weighting functions presented here
do not consider the size of the pattern, nor the numbers of defined and undefined
constants. Considering more complicated weighting functions may be necessary
for formal libraries with significantly different logics.

Based on the weighting functions two scoring functions are defined.
Let $c_1$ be a constant from library $lib_1$ and $c_2$ a constant from library $lib_2$. Let $P = \lbrace p_1, \ldots, p_k \rbrace$ be the set of patterns $c_1$ and $c_2$ have in common. Then:
\begin{eqnarray*}
 score_0(c1,c2) &=& \sum\limits_{i=1}^k w_0(lib_1,p_i) * w_0(lib_2,p_i)\\
 score_1(c1,c2) &=& \sum\limits_{i=1}^k w_1(lib_1,p_i) * w_1(lib_2,p_i)\\
\end{eqnarray*}
 In order to account for the fact that constants with a high number of associated patterns are more likely to have common patterns with unrelated constants, we further modify $score_1$. Let $n_1$ be the number of patterns associated to $c_1$ and $n_2$ be the number of patterns associated to $c_2$. We define a third similarity scoring function by:
     \[ score_2(c1,c2) = \frac
     {\sum\limits_{i=1}^k w_1(lib_1,p_i) * w_1(lib_2,p_i)}
      {log ( 2  + n_1 * n_2)}
      \]

\subsection{Iterative approach}
In our initial experiments, a direct computation of the $score_i$ functions for all
constants in two libraries after an initial number of correct pairs would find
incorrect pairs (false positive matches). Such pairs can be quickly eliminated if the information coming
from the first successful matches is propagated further. In order to do this,
we propose an iterative approach (presented schematically in Fig.~\ref{fig:iterative}):
\begin{figure}[ht]
\centering
\begin{tikzpicture}[node distance = 2cm, auto]
  % Place nodes
   
   \node [] (export1) {};
   \node [below of=export1,node distance=1cm] (export2) {};
   \node [cloud,right of=export1,  node distance=2cm] 
   (theorems1) {theorems 1};
   \node [cloud, right of=theorems1,  node distance=3cm] 
   (patterns1){patterns 1};
   \node [cloud, right of=patterns1, yshift = -14, node distance=4cm] 
   (score){ranked constant pairs};
   \node [cloud, right of= export2,node distance=2cm] (theorems2) {theorems 2};
   \node [cloud, right of=theorems2,node distance=3cm] 
   (patterns2){patterns 2};
   \node [ right of=score,  node distance=3cm] (result) {};
   % Draw edges
   \draw [-to,black,thick] (export1) -- 
   node[name=up1]{0} (theorems1);
   \draw [-to,black,thick] (export2) -- 
   node[name=down1]{0} (theorems2);
   \draw [-to,black,thick] (theorems1) -- 
   node[name=up1]{1} (patterns1);
   \draw [-to,black,thick] (theorems2) -- 
   node[name=down1]{1} (patterns2);
   \draw [-to,black,thick] (patterns1) -- node[name=up2]{2}(score);
   \draw [-to,black,thick] (patterns2) -- node[name=down2]{2}(score);
   \draw [-to,black,thick] (score) to [out=210,in=340] node[name=up3]{3} (theorems2);
   \draw [-to,black,thick] (score) to [out=150,in=20] node[name=down3]{3} (theorems1);
   \draw [-to,black,thick] (score) -- node[name=down4]{4}(result);
\end{tikzpicture}
\caption{Graphical representation of the iterative procedure}
\label{fig:iterative}
\end{figure}
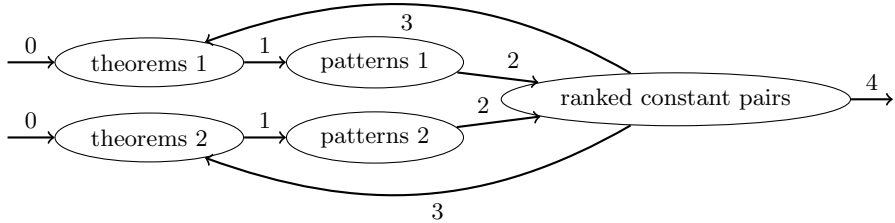

The iterative approach returns a sorted list of pairs of constants and a sorted list of pair of types from two libraries by following this steps:
 \begin{itemize}
  \item[0.] Export theorems from a library as well as constants with their types and parse them.
  \item[1.] Normalize theorems and create theorem patterns, constant patterns and type patterns according to the current defined constants and types.
  \item[2.] Score every pair of constants.
  \item[3.] Take the highest ranked pair of constants $(c_1,c_2)$. Check if their type matches, if not take the next one and so on. When their type matches, rewrite all the theorems inside $lib_1$ with the substitution $c_1 \rightarrow d$ 
  and all the  theorems inside $lib_2$ with the substitution $c_2 \rightarrow d$, where $d$ is a fresh defined constant. Then, get the derived pairs of types from the pair of constant and substitute every pair member with the same fresh defined type as for the other member.
  \item[4.] Return the pairs of constants and the pairs of types, in the order they were created, when you have reached the number of iteration desired.
 \end{itemize}
The single-pass approach is defined by doing only one iteration, where the list of pairs of constants are returned ranked by their score. A type check performed after a single-pass can discard a number of wrong matches efficiently.
 
 In the presented approach, we assume that the constants and types inside one library are all
 different, which we tried to ensure by the initial normalization. Thus, we will
 not match constant from the same library. Furthermore, if a constant is matched, then it
 can no longer be matched again and the same reasoning applies for types. This first statement
 will turn out not to be true for a few constants in Section \ref{s:experiment}.

The complexity of the iterative approach is obviously larger than that of the single-pass approach.
On an IntelM 2.66GHz CPU,
the single-pass approach between \holfour and \hollight with $score_2$ and $norm_2$ takes 6
minutes to complete. The main reason is that it has to compare the patterns of all possible
pairs of constants (about two million). Thus, the bottleneck is the time taken by the comparison
function which intersects the set of patterns associated with each constant and scores the
resulting set. However, the iterative method can use the first pass to remove pairs of constants
that have no common patterns. This reduces the number of possible matches to ten thousand. As
a consequence, it takes only 3 minutes more to do 100 iterations.

\section{Experiments}\label{s:experiment}
In order to verify the correctness of our approach we first investigate the most common patterns
and shapes of theorems in each of the three formal libraries and then we look at the results
of the matching constants across libraries. The data given by these experiments is available at \url{http://cl-informatik.uibk.ac.at/users/tgauthier/matching/}.

\subsection{Single library results}
Tables \ref{tab:prop1} and \ref{tab:prop2} show the most common properties when
applying the standard normalization $norm_1$ of a single constant and of two constants
respectively in the three considered proof assistant libraries. 
The tables are sorted with respect to the total number of different constants in
the theorems from which the patterns are derived. In Table~\ref{tab:prop1}, $Inj$ stands
for injectivity, $Asso$ for associativity and $Comm$ for commutativity. In Table~\ref{tab:prop2},
the pattern $Class$ and $Inv$ are defined by $Class \ (c_0, c_1) = c_0\ c_1$, $Inv(c_0,c_1) = \forall x_0.\ c_0\ (c_1 \ x_0) = x_0$.
  
 \begin{table*}[t]
 \centering\ra{1.3}
 \begin{tabular}{@{}lccclccclcc@{}}
 \toprule 
 \multicolumn{3}{c}{\hollight} & \phantom{abc} & 
 \multicolumn{3}{c}{\holfour}  & \phantom{abc} & 
 \multicolumn{3}{c}{\isabellehol} \\
 \cmidrule{1-3} \cmidrule{5-7} \cmidrule{9-11}
  Pattern & Consts & Thms && 
  Pattern & Consts & Thms &&
  Pattern & Consts & Thms \\
 \midrule
    Inj & 37 & 37 && 
    Inj & 54 & 68 && 
    Inj & 83 & 137\\
    Asso & 32 & 36 && 
    Asso & 50 & 65 && 
    App & 17 & 18\\
    Comm & 25 & 44 && 
    Comm & 40 & 48 && 
    Inj1 & 16 & 16\\
    Refl & 22 & 22 && 
    Trans & 32 & 33 && 
    Comm & 14 & 51\\
    Lcomm & 19 & 20 && 
    Refl & 23 & 23 && 
    Inj2 & 12 & 35\\
    Idempo & 12 & 12 && 
    Idempo & 20 & 15 && 
    App2 & 11 & 12\\
 \bottomrule
 \end{tabular}
   \caption{Most frequent properties of one constant} 
   \label{tab:prop1}
 \end{table*}  
 
 \begin{table*}[t] 
 \centering\ra{1.3}
 \begin{tabular}{@{}lccclccclcc@{}}
 \toprule 
 \multicolumn{3}{c}{\hollight} & \phantom{abc} & 
 \multicolumn{3}{c}{\holfour}  & \phantom{abc} & 
 \multicolumn{3}{c}{\isabellehol} \\
 \cmidrule{1-3} \cmidrule{5-7} \cmidrule{9-11}
  Pattern & Consts & Thms && 
  Pattern & Consts & Thms &&
  Pattern & Consts & Thms \\
 \midrule
    Class & 71 & 87 && 
    Inv & 131 & 89 && 
    Class & 188 & 642\\
    Inv & 64 & 34 && 
    Neutr & 64 & 55 && 
    Inv & 114 & 75\\
    Imp & 52 & 76 && 
    Class & 63 & 70 && 
    Equal & 58 & 40\\
 \bottomrule
 \end{tabular}
   \caption{Most frequent properties of two constants} 
   \label{tab:prop2}
 \end{table*}

As expected, \hollight and \holfour show the most similar results and injectivity is the most frequent property. Commutativity and associativity are also very common, and their associated constants are used to apply $norm_2$ as stated in Section \ref{s:classify}.

The common patterns immediately show constants defined to be equivalent to
the defined equality in each of the libraries, through an extensional definition. There
is one such constant in \holfour, one in \hollight and three in \isabellehol. In order to
avoid missing or duplicate patterns we mapped all these constants to equality manually.

Furthermore, in Table~\ref{tab:prop2}, the third row of the \isabellehol column shows 40 equalities between two different constants that were created during the normalization. We have also found 10 such equalities in \holfour and 1 in \hollight. Often a constant with a less general type can be replaced by the other, but without type-class information
in \isabellehol we decided not to do such replacements in general.

\subsection{Cross-library results}
The way we analyze the quality of the matching, is by looking at the number of correct matches
of types and constants between the libraries, in particular we consider the occurrence of the
first incorrect match, also called \emph{false positive} below. It is very hard to spot same
concepts in two large libraries, therefore a manual evaluation of the false negatives
(pairs that could be mapped but are not) is a very hard task and requires
the knowledge of the whole libraries.
% We do not know of such false negative
%pairs, but it is likely that some exist.

In the first three experiments, we test how much normalization, scoring, iteration and types contribute
to better matches. This will be used to choose the best parameters for matching constants and
types between each pair of provers.

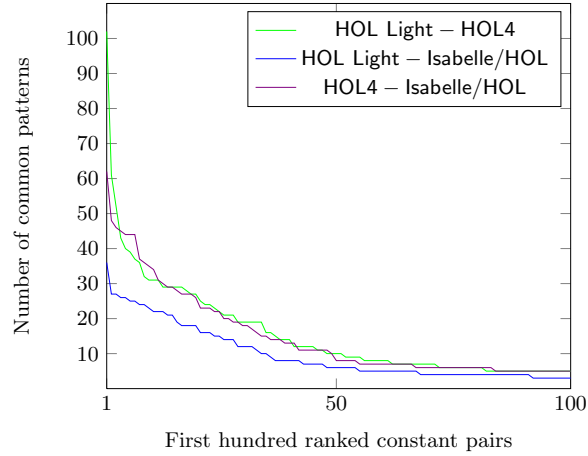
\begin{figure}[htb]
\centering
\begin{tikzpicture}[scale=0.9]
  \begin{axis}[ 
            xmin=1, xmax=100,
            ymin=0, ymax=110,
            xtick={1,50,100},
            ytick={10,20,30,40,50,60,70,80,90,100},
            xlabel={First hundred ranked constant pairs},
            ylabel={Number of common patterns}
          ]  
   \addplot [green] coordinates{
 (1,102.)(2,61.)(3,52.)(4,43.)(5,40.)(6,39.)(7,37.)(8,36.)(9,32.)(10,31.)(11,31.)(12,31.)(13,29.)(14,29.)(15,29.)(16,29.)(17,29.)(18,28.)(19,27.)(20,27.)(21,25.)(22,24.)(23,24.)(24,23.)(25,22.)(26,21.)(27,21.)(28,21.)(29,19.)(30,19.)(31,19.)(32,19.)(33,19.)(34,19.)(35,16.)(36,16.)(37,15.)(38,14.)(39,14.)(40,14.)(41,12.)(42,12.)(43,12.)(44,12.)(45,12.)(46,11.)(47,11.)(48,10.)(49,10.)(50,10.)(51,10.)(52,9.)(53,9.)(54,9.)(55,9.)(56,8.)(57,8.)(58,8.)(59,8.)(60,8.)(61,8.)(62,7.)(63,7.)(64,7.)(65,7.)(66,7.)(67,7.)(68,7.)(69,7.)(70,7.)(71,7.)(72,6.)(73,6.)(74,6.)(75,6.)(76,6.)(77,6.)(78,6.)(79,6.)(80,6.)(81,6.)(82,5.)(83,5.)(84,5.)(85,5.)(86,5.)(87,5.)(88,5.)(89,5.)(90,5.)(91,5.)(92,5.)(93,5.)(94,5.)(95,5.)(96,5.)(97,5.)(98,5.)(99,5.)(100,5.)
   };
   
   \addplot [blue] coordinates{
(1,36.)(2,27.)(3,27.)(4,26.)(5,26.)(6,25.)(7,25.)(8,24.)(9,24.)(10,23.)(11,22.)(12,22.)(13,22.)(14,21.)(15,21.)(16,19.)(17,18.)(18,18.)(19,18.)(20,18.)(21,16.)(22,16.)(23,16.)(24,15.)(25,15.)(26,14.)(27,14.)(28,14.)(29,12.)(30,12.)(31,12.)(32,12.)(33,11.)(34,10.)(35,10.)(36,9.)(37,8.)(38,8.)(39,8.)(40,8.)(41,8.)(42,8.)(43,7.)(44,7.)(45,7.)(46,7.)(47,7.)(48,6.)(49,6.)(50,6.)(51,6.)(52,6.)(53,6.)(54,6.)(55,5.)(56,5.)(57,5.)(58,5.)(59,5.)(60,5.)(61,5.)(62,5.)(63,5.)(64,5.)(65,5.)(66,5.)(67,5.)(68,4.)(69,4.)(70,4.)(71,4.)(72,4.)(73,4.)(74,4.)(75,4.)(76,4.)(77,4.)(78,4.)(79,4.)(80,4.)(81,4.)(82,4.)(83,4.)(84,4.)(85,4.)(86,4.)(87,4.)(88,4.)(89,4.)(90,4.)(91,4.)(92,3.)(93,3.)(94,3.)(95,3.)(96,3.)(97,3.)(98,3.)(99,3.)(100,3.) 
    };       

\addplot [violet] coordinates{      
   (1,62.)(2,48.)(3,46.)(4,45.)(5,44.)(6,44.)(7,44.)(8,37.)(9,36.)(10,35.)(11,34.)(12,31.)(13,30.)(14,29.)(15,29.)(16,28.)(17,27.)(18,27.)(19,27.)(20,26.)(21,23.)(22,23.)(23,23.)(24,22.)(25,22.)(26,20.)(27,20.)(28,19.)(29,19.)(30,18.)(31,18.)(32,17.)(33,16.)(34,15.)(35,15.)(36,14.)(37,14.)(38,14.)(39,13.)(40,13.)(41,13.)(42,11.)(43,11.)(44,11.)(45,11.)(46,11.)(47,11.)(48,11.)(49,10.)(50,8.)(51,8.)(52,8.)(53,8.)(54,8.)(55,7.)(56,7.)(57,7.)(58,7.)(59,7.)(60,7.)(61,7.)(62,7.)(63,7.)(64,7.)(65,7.)(66,7.)(67,6.)(68,6.)(69,6.)(70,6.)(71,6.)(72,6.)(73,6.)(74,6.)(75,6.)(76,6.)(77,6.)(78,6.)(79,6.)(80,6.)(81,6.)(82,6.)(83,6.)(84,5.)(85,5.)(86,5.)(87,5.)(88,5.)(89,5.)(90,5.)(91,5.)(92,5.)(93,5.)(94,5.)(95,5.)(96,5.)(97,5.)(98,5.)(99,5.)(100,5.)
   };
    \legend{$\hollight-\holfour$,
                $\hollight-\isabellehol$,
                $\holfour-\isabellehol$}
   \end{axis}
  \end{tikzpicture}
      \caption{Number of patterns by constant pairs in different provers} 
      \label{fig:patternprover}
    \end{figure}

The first experiment (Fig.~\ref{fig:patternprover}) evaluates the similarity of the libraries.
We match the provers using the (a-priori) strongest normalization ($norm_2$) with a single-pass
approach with no types. In this setting, the
constant with the most similar properties is $0$ between \hollight and \holfour, and between
\holfour and \isabellehol. And it is $\emptyset$ between \hollight and \isabellehol.
Form this perspective, the most similar pairs of provers are in decreasing order (\hollight,\holfour), (\holfour,\isabellehol) and (\hollight-\isabellehol). We test the four other parameters relative to the pair of
provers (\hollight, \holfour) as we should have most common patterns to work with.

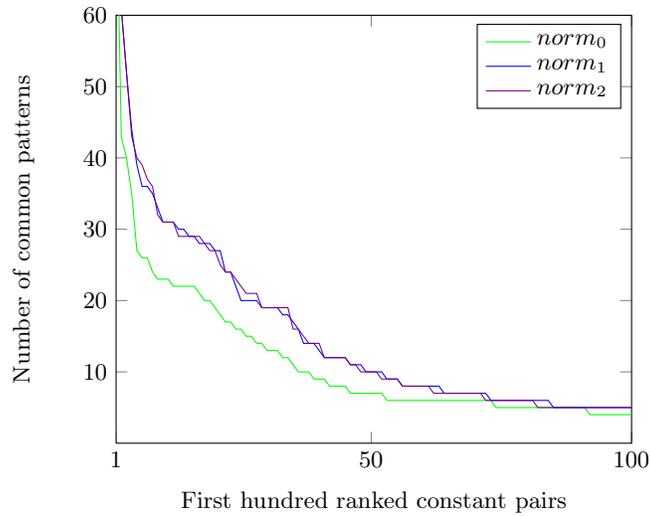
\begin{figure}[H] 
\centering
\begin{tikzpicture}
  \begin{axis}[ 
            xmin=1, xmax=100,
            ymin=0, ymax=60,
            xtick={1,50,100},
            ytick={10,20,30,40,50,60},
            xlabel={First hundred ranked constant pairs},
            ylabel={Number of common patterns}
          ]  
\addplot [green] coordinates{      
(1,81.)(2,43.)(3,40.)(4,35.)(5,27.)(6,26.)(7,26.)(8,24.)(9,23.)(10,23.)(11,23.)(12,22.)(13,22.)(14,22.)(15,22.)(16,22.)(17,21.)(18,20.)(19,20.)(20,19.)(21,18.)(22,17.)(23,17.)(24,16.)(25,16.)(26,15.)(27,15.)(28,14.)(29,14.)(30,13.)(31,13.)(32,13.)(33,12.)(34,12.)(35,11.)(36,10.)(37,10.)(38,10.)(39,9.)(40,9.)(41,9.)(42,8.)(43,8.)(44,8.)(45,8.)(46,7.)(47,7.)(48,7.)(49,7.)(50,7.)(51,7.)(52,7.)(53,6.)(54,6.)(55,6.)(56,6.)(57,6.)(58,6.)(59,6.)(60,6.)(61,6.)(62,6.)(63,6.)(64,6.)(65,6.)(66,6.)(67,6.)(68,6.)(69,6.)(70,6.)(71,6.)(72,6.)(73,6.)(74,5.)(75,5.)(76,5.)(77,5.)(78,5.)(79,5.)(80,5.)(81,5.)(82,5.)(83,5.)(84,5.)(85,5.)(86,5.)(87,5.)(88,5.)(89,5.)(90,5.)(91,5.)(92,4.)(93,4.)(94,4.)(95,4.)(96,4.)(97,4.)(98,4.)(99,4.)(100,4.)  };
   
   \addplot [blue] coordinates{
(1,105.)(2,61.)(3,52.)(4,44.)(5,39.)(6,36.)(7,36.)(8,35.)(9,33.)(10,31.)(11,31.)(12,31.)(13,30.)(14,30.)(15,29.)(16,29.)(17,28.)(18,28.)(19,28.)(20,27.)(21,27.)(22,24.)(23,24.)(24,22.)(25,20.)(26,20.)(27,20.)(28,20.)(29,19.)(30,19.)(31,19.)(32,19.)(33,18.)(34,18.)(35,17.)(36,16.)(37,14.)(38,14.)(39,14.)(40,13.)(41,12.)(42,12.)(43,12.)(44,12.)(45,12.)(46,11.)(47,11.)(48,11.)(49,10.)(50,10.)(51,10.)(52,10.)(53,9.)(54,9.)(55,9.)(56,8.)(57,8.)(58,8.)(59,8.)(60,8.)(61,8.)(62,8.)(63,8.)(64,7.)(65,7.)(66,7.)(67,7.)(68,7.)(69,7.)(70,7.)(71,7.)(72,7.)(73,6.)(74,6.)(75,6.)(76,6.)(77,6.)(78,6.)(79,6.)(80,6.)(81,6.)(82,6.)(83,6.)(84,6.)(85,5.)(86,5.)(87,5.)(88,5.)(89,5.)(90,5.)(91,5.)(92,5.)(93,5.)(94,5.)(95,5.)(96,5.)(97,5.)(98,5.)(99,5.)(100,5.)
    };       

   \addplot [violet] coordinates{
(1,102.)(2,61.)(3,52.)(4,43.)(5,40.)(6,39.)(7,37.)(8,36.)(9,32.)(10,31.)(11,31.)(12,31.)(13,29.)(14,29.)(15,29.)(16,29.)(17,29.)(18,28.)(19,27.)(20,27.)(21,25.)(22,24.)(23,24.)(24,23.)(25,22.)(26,21.)(27,21.)(28,21.)(29,19.)(30,19.)(31,19.)(32,19.)(33,19.)(34,19.)(35,16.)(36,16.)(37,15.)(38,14.)(39,14.)(40,14.)(41,12.)(42,12.)(43,12.)(44,12.)(45,12.)(46,11.)(47,11.)(48,10.)(49,10.)(50,10.)(51,10.)(52,9.)(53,9.)(54,9.)(55,9.)(56,8.)(57,8.)(58,8.)(59,8.)(60,8.)(61,8.)(62,7.)(63,7.)(64,7.)(65,7.)(66,7.)(67,7.)(68,7.)(69,7.)(70,7.)(71,7.)(72,6.)(73,6.)(74,6.)(75,6.)(76,6.)(77,6.)(78,6.)(79,6.)(80,6.)(81,6.)(82,5.)(83,5.)(84,5.)(85,5.)(86,5.)(87,5.)(88,5.)(89,5.)(90,5.)(91,5.)(92,5.)(93,5.)(94,5.)(95,5.)(96,5.)(97,5.)(98,5.)(99,5.)(100,5.)
   };

        \legend{$norm_0$,$norm_1$,$norm_2$}
        \end{axis}
        \end{tikzpicture}
      \caption{The normalization effect} 
    \label{fig:normalization}
    \end{figure}

The second experiment (Fig.~\ref{fig:normalization}) is meant to evaluate the efficiency of normalization on the number of patterns. It is also run as a single-pass with no types.
We observe an increase in number of patterns from $norm_0$ and $norm_1$ which is mostly due to the splitting of disjunctions. Moreover, the difference between $norm_2$ and $norm_1$ is negligible, which means that associative and commutative constants are used in almost the same way across the two libraries.
In the following experiments we will only use $norm_2$ assuming it is the strongest normalization also in the other scenarios.

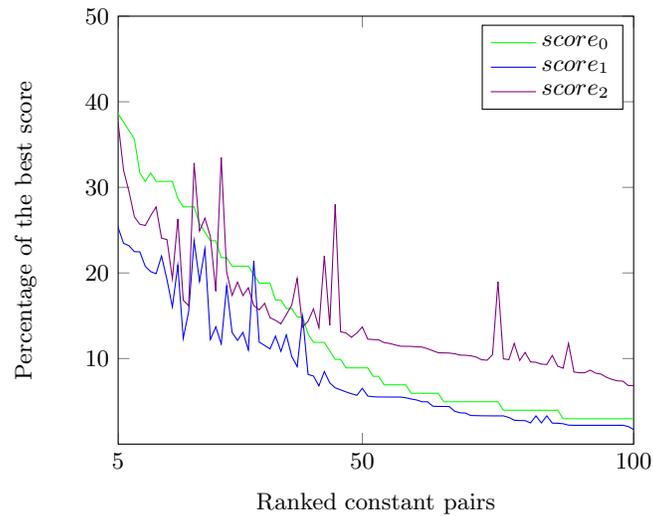
\begin{figure}[H]
           \centering
           \begin{tikzpicture}
             \begin{axis}[ 
                       xmin=5, xmax=100,
                       ymin=0, ymax=50,
                       xtick={5,50,100},
                       ytick={10,20,30,40,50,60,70,80,90,100},
                       xlabel={Ranked constant pairs},
                       ylabel={Percentage of the best score}
                     ]  
            \addplot [green,yscale=0.99] coordinates{
            (1,102.)(2,61.)(3,51.)(4,43.)(5,39.)(6,38.)(7,37.)(8,36.)(9,32.)(10,31.)(11,32.)(12,31.)(13,31.)(14,31.)(15,31.)(16,29.)(17,28.)(18,28.)(19,28.)(20,26.)(21,25.)(22,24.)(23,24.)(24,22.)(25,22.)(26,21.)(27,21.)(28,21.)(29,21.)(30,20.)(31,19.)(32,19.)(33,19.)(34,17.)(35,17.)(36,16.)(37,16.)(38,15.)(39,15.)(40,13.)(41,12.)(42,12.)(43,12.)(44,11.)(45,10.)(46,10.)(47,9.)(48,9.)(49,9.)(50,9.)(51,9.)(52,8.)(53,8.)(54,7.)(55,7.)(56,7.)(57,7.)(58,7.)(59,6.)(60,6.)(61,6.)(62,6.)(63,6.)(64,6.)(65,5.)(66,5.)(67,5.)(68,5.)(69,5.)(70,5.)(71,5.)(72,5.)(73,5.)(74,5.)(75,5.)(76,4.)(77,4.)(78,4.)(79,4.)(80,4.)(81,4.)(82,4.)(83,4.)(84,4.)(85,4.)(86,4.)(87,3.)(88,3.)(89,3.)(90,3.)(91,3.)(92,3.)(93,3.)(94,3.)(95,3.)(96,3.)(97,3.)(98,3.)(99,3.)(100,3.)
              };
              \addplot[blue,yscale=1.1] coordinates{
           (1,90.9926619133)(2,45.0665178165)(3,47.4478021978)(4,25.6306337084)(5,23.)(6,21.3285940546)(7,21.0721611722)(8,20.4409722222)(9,20.4213203463)(10,18.8853252412)(11,18.3220238095)(12,18.1027777778)(13,19.9760259371)(14,17.4741017316)(15,14.5824423534)(16,18.9969549486)(17,11.2237515519)(18,14.0897789453)(19,21.5976911977)(20,17.3011349489)(21,20.7359260021)(22,11.1166666667)(23,12.4544144173)(24,10.6518903145)(25,16.8)(26,11.856960034)(27,11.012221723)(28,11.8645768415)(29,10.0170049858)(30,19.4594642857)(31,10.8336996337)(32,10.4804509335)(33,10.1091838334)(34,11.4741574039)(35,9.85096292596)(36,11.5955555556)(37,9.3)(38,8.23249389499)(39,13.5315934066)(40,7.40436507937)(41,7.22916666667)(42,6.19203296703)(43,7.7037037037)(44,6.50527777778)(45,5.97916666667)(46,5.76351495726)(47,5.54341491841)(48,5.35151515152)(49,5.18484035028)(50,5.9287037037)(51,5.10416666667)(52,5.03333333333)(53,5.00699300699)(54,5.00649350649)(55,5.)(56,5.)(57,5.)(58,4.9273633157)(59,4.79166666667)(60,4.69166666667)(61,4.5)(62,4.5)(63,4.025)(64,4.)(65,4.)(66,4.)(67,3.5)(68,3.33406700905)(69,3.3)(70,3.03703703704)(71,3.02777777778)(72,3.00925925926)(73,3.00699300699)(74,3.)(75,3.)(76,3.)(77,2.83333333333)(78,2.54166666667)(79,2.5)(80,2.5)(81,2.25)(82,3.)(83,2.25)(84,3.)(85,2.23333333333)(86,2.21742424242)(87,2.16666666667)(88,2.01770482742)(89,2.)(90,2.)(91,2.)(92,2.)(93,2.)(94,2.)(95,2.)(96,2.)(97,2.)(98,2.)(99,1.87435897436)(100,1.54166666667)
               };       
               \addplot[violet,yscale=17] 
                   coordinates{      
                  (1,5.74497045271)(2,3.09023058389)(3,3.09403530554)(4,2.26730503196)(5,2.20786754643)(6,1.88275636047)(7,1.7359860965)(8,1.56189403577)(9,1.5109727682)(10,1.50267091033)(11,1.57034100205)(12,1.63028859736)(13,1.41582111927)(14,1.40565925012)(15,1.1363715263)(16,1.54722676687)(17,0.98901395994)(18,0.950729287027)(19,1.93125900959)(20,1.46478184609)(21,1.55309888589)(22,1.42679590916)(23,1.05098834067)(24,1.97019937381)(25,1.18179331135)(26,1.02208997459)(27,1.11308726246)(28,1.02078224046)(29,1.07429161338)(30,0.951753505443)(31,0.922627966688)(32,0.967971869233)(33,0.868588963807)(34,0.850976855283)(35,0.826357830197)(36,0.891062839338)(37,0.955780269831)(38,1.1388074483)(39,0.80419517695)(40,0.844820545103)(41,0.92930879808)(42,0.80000344706)(43,1.29433176619)(44,0.815801882925)(45,1.64727740517)(46,0.772242479499)(47,0.76385615091)(48,0.734464491909)(49,0.763889564258)(50,0.805520500218)(51,0.721347520444)(52,0.718668347241)(53,0.716801552774)(54,0.697955438512)(55,0.691952512522)(56,0.68223598087)(57,0.672745187386)(58,0.67227626055)(59,0.671553852336)(60,0.669121982769)(61,0.667061112285)(62,0.655837368745)(63,0.641802384774)(64,0.627969059722)(65,0.626777922756)(66,0.625854553674)(67,0.62133493456)(68,0.611586533987)(69,0.610570566348)(70,0.607085609083)(71,0.598174470506)(72,0.579683837241)(73,0.576714314103)(74,0.615254292072)(75,1.1162212531)(76,0.586016157879)(77,0.580435051613)(78,0.691952512522)(79,0.575149991929)(80,0.629964390439)(81,0.565477425359)(82,0.562009099957)(83,0.549552779936)(84,0.547260672522)(85,0.609953632941)(86,0.535819794744)(87,0.521614272193)(88,0.692237266643)(89,0.49527631739)(90,0.49049212362)(91,0.49049212362)(92,0.508483422712)(93,0.48806787092)(94,0.480898346963)(95,0.459062122249)(96,0.445077600927)(97,0.434682938506)(98,0.433083995117)(99,0.402429604382)(100,0.402429604382)
                  };
                  
                  \legend{$score_0$,$score_1$,$score_2$}
                  \end{axis}
              
               \end{tikzpicture}
                 \caption{Effect of different scoring functions on the iterative approach} 
                 \label{fig:iterativeeffect}
               \end{figure}

We next evaluate the scoring functions, the contribution of iterations, and of the use of type information.
%The results of this experiment are shown in two tables and a plot.
Table~\ref{tab:wrongmatch} shows the effect of
iterative method and scoring function on the occurrence of the first wrong match (false positive). It has been inspected manually.
Fig.~\ref{fig:iterativeeffect}, shows the positive effects of the iterative effect on the $score_1$ and $score_2$ curves.
Some patterns are ranked higher after an iteration, as they become more scarce. The iterative method also has an
opposite effect that is not directly visible in the figure: the score of pairs of constants diminishes by removing false pattern matches.
 Table~\ref{tab:wrongtype} shows how type information contributes to matches. Types do help,
but become less important with better scoring functions combined with the iterative approach.

\begin{table*}
\centering\ra{1.3}
\begin{tabular}{@{}lcccccc@{}}
\toprule & \phantom{abc} & 
          \multicolumn{1}{c}{$score_0$} & \phantom{abc} & 
          \multicolumn{1}{c}{$score_1$}  & \phantom{abc} & 
          \multicolumn{1}{c}{$score_2$} \\
          \midrule
             Single-pass && 39 && 69 && 88  \\
             Iterative && 49 && 68 && 113 \\
     
          \bottomrule
\end{tabular}
\caption{Rank of the first wrong match for (\hollight, \holfour)} 
\label{tab:wrongmatch}
\end{table*}     
\begin{table*}
          \centering\ra{1.3}
          \begin{tabular}{@{}lcccccc@{}}
          \toprule 
          
          & \phantom{abc} & 
          \multicolumn{1}{c}{$score_0$} & \phantom{abc} & 
          \multicolumn{1}{c}{$score_1$}  & \phantom{abc} & 
          \multicolumn{1}{c}{$score_2$} \\
          \midrule
             Single-pass && 31 && 19 && 21 \\
             Iterative && 224 && 18 && 6 \\
          \bottomrule
          \end{tabular}
            \caption{Number of pairs of constants discarded, due to type matching 
            \label{tab:wrongtype}
            } 
          \end{table*}             
          
The last experiment is run with the best parameters found by the previous experiments, namely $norm_2$, $score_2$ and
the iterative approach with types. Three numbers are presented in each cell of Tables~\ref{tab:final1} and \ref{tab:final2}. The first one is the number of correct matches obtained before the first error. The second one is number of correct matches we have found. In the case of constants, the third one is the number of matches we have manually checked.
We stop at a point where a previously found error propagates. In the case of types, the third number is the rank of the last correct match. As seen previously, the best results come from comparing the \holfour and \hollight libraries, where we have verified 177 constant matches and 16 type matches.          
% As the checking was done manually, it is subject to errors raising or reducing the number presented by a small margin.          

 \begin{table}[H]
 \centering\ra{1.3}
 \begin{tabular}{@{}cccccccccc@{}}
 \toprule 
 \multicolumn{2}{c}{\hollight-\holfour} & \phantom{abc} & 
 \multicolumn{2}{c}{\holfour-\isabellehol}  & \phantom{abc} & 
 \multicolumn{2}{c}{\hollight-\isabellehol} \\
 \cmidrule{1-2} \cmidrule{4-5} \cmidrule{7-8}
  112 & 177/203 &&
  65 & 109/131 && 
  55 & 78/98 \\
 \bottomrule
 \end{tabular}
   \caption{Number of constants accurately matched} 
   \label{tab:final1}
 \end{table}  

 \begin{table}[H]
  \centering\ra{1.3}
  \begin{tabular}{@{}cccccccccc@{}}
  \toprule 
  \multicolumn{2}{c}{\hollight-\holfour} & \phantom{abc} & 
  \multicolumn{2}{c}{\holfour-\isabellehol}  & \phantom{abc} & 
  \multicolumn{2}{c}{\hollight-\isabellehol} \\
  \cmidrule{1-2} \cmidrule{4-5} \cmidrule{7-8}
   11 & 16/22 &&
   8 & 11/17 && 
   6 & 7/13 \\
  \bottomrule
  \end{tabular}
    \caption{Number of types accurately matched} 
    \label{tab:final2}
  \end{table}

\section{Conclusion}\label{s:concl}
We have investigated the formal mathematical libraries of \hollight, \holfour and \isabellehol
searching for common types and constants. We defined a concept of patterns that capture abstract properties
of constants and types and normalization on theorems that allow for efficient computation of
such patterns. The practical evaluation of the approach on the libraries let us find hundreds
of pairs of common patterns, with a high accuracy.

Formal mathematical libraries contain many instances of the same algebraic structures. Such
instances have many same properties therefore their matching is non-trivial. Our proposed
approach can match such instances correctly, because of patterns that link such concepts to
other concepts. For example integers and matrices are instances of the algebraic structure 
ring. However each of the libraries we analyzed contains a theorem that states that each integers
is equal to a natural number or
its negation. A pattern derived from this fact, together with many other patterns that are unique
to integers match them across libraries correctly.

%Therefore, the design of a concept matching algorithm was tested at a big scale. The exporting phase was performed in a way that enable us to easily parse the result into our main project. The creation of patterns and the normalization were also defined to make theorems from each library less dependent on their specific structure. From that, we were able to extract properties of constants inside a library and see if they were occurring in other libraries.
%As a result, we designed an improved normalization using associative and commutative constants and rewrote the similarity scoring function to take into account the frequency of a pattern.
%We then layout a iterative approach so that the matching algorithm can learn form its previous matches. Finally, we added type information as a filter to reduce the number of errors.
%All in all, the improved normalization was shown to have little to no efficiency. The typing information and the iterative approach contributed to better results and some experiments exhibited their redundant and complementary role. In the end, most of the latest progress was achieved by improving the scoring function. In average, we have managed to match accurately 121 constants and 11 types between each pair of provers, showing that our method is not specifically designed for a fixed pair of provers.

%\subsection{Future work}
The work gives many correct matches between concepts that can be directly
used in translations between proof assistants. In particular \textsc{HOL/Import} would immediately
benefit from mapping the \hollight types and constants to their \isabellehol counterparts
allowing for further merging of the results.

The approach has been tested on three provers based on higher-order logic. In principle
the properties of the standard mathematical concepts defined in many other logics are the
same, however it remains to be seen how smoothly does the approach extend to provers based
on different logics.

In order to further decrease the number of false positive matches, more weighting and
scoring functions could be considered. One could imagine functions that take into account
the length of formulas, and numbers of mapped concepts per pattern. Similarly, the scoring
functions could penalize pairs of constants with only one pattern in common (these have been
the first false positives in all our experiments). Further, the equalities between constants created during normalization could be used for further rewriting of theorems into normal forms.
Other ideas include normalizing relatively to distributive constants and trying weaker kind
of matching for example on atoms or subterms.

Building a set of basic mathematical concepts together with their
foundational properties has been on the MKM wish-list for a long time. It remains
to be seen if a set of common concepts across proof assistant libraries can be extended
by minimal required properties to automatically build such ``interface theories'', and
if automatically found larger sets of theories can complement the high-quality interface
theories built in the MKM community.

% Apply translation to proof advice~\cite{ckju-cicm13}.

\section*{Acknowledgments}

We would like to thank Josef Urban for his comments on the previous version of this paper.\\
This work has been supported by the Austrian Science Fund (FWF): P26201.

\bibliographystyle{plain}
\bibliography{biblio}

\end{document}